\documentclass[10pt]{iopart}

\usepackage[backend=biber, style=phys]{biblatex}  
\expandafter\let\csname equation*\endcsname\relax
\expandafter\let\csname endequation*\endcsname\relax
\usepackage{amsmath, amsbsy}
\usepackage{graphicx}
\usepackage[unicode=true, bookmarks=true, bookmarksnumbered=true, bookmarksopen=true, bookmarksopenlevel=1, breaklinks=false,pdfborder={0 0 0}, backref=false, colorlinks=true]{hyperref}
\hypersetup{citecolor=blue,linkcolor=blue}
\usepackage{breakurl}

\usepackage[dvipsnames]{xcolor} 
\usepackage{lipsum,multicol} 

\usepackage [english]{babel}
\usepackage [autostyle, english = american]{csquotes}
\MakeOuterQuote{"}

\usepackage{subcaption}

\usepackage{graphicx}

\usepackage{cleveref}

\usepackage[final,commandnameprefix=always]{changes}

\usepackage{siunitx}

\begin{document}



\title{Determining the Validity of Tokamak Perturbed Equilibrium Modeling Using Nonlinear Equilibria}

\author{
J. Halpern$^1$,
N.C. Logan$^1$,
E. Paul$^1$,
C. Paz-Soldan$^1$
}
\address{
$^1$Columbia University, New York, New York 10027, USA
 }
 
\vspace{10pt}
\begin{indented}
\item[]\today
\end{indented}

\begin{abstract}

The prediction of perturbed equilibrium models for tokamaks with small non-axisymmetric fields is strongly dependent on which reference frame for axisymmetry is assumed. This assumption directly affects the applied field spectrum on the plasma, with the incorrect choice resulting in an incorrect prediction of the plasma response. We use fully 3D equilibria generated by VMEC to determine the correct reference frame when calculating error fields in perturbed equilibrium codes subject to $n=1$ misalignments in several coil sets. We analyze a case of independently offset toroidal field coils and central solenoid/poloidal field coils in the SPARC tokamak and find that the appropriate reference frame can be well approximated by the centroid of the toroidal field coil set. We also consider the case of NSTX-U with an independent centerpost consisting of the inner legs of the toroidal field coils and central solenoid, and find that the reference frame can be well approximated by the radial location of the toroidal field coil inboard legs at the midplane. We determine the correct frame by analyzing the shifted magnetic axis in the nonlinear equilibria, and use our findings to generalize to 3D fields from misalignments which modify the reference frame and those which impact the plasma response. We also analyze the magnetic field line displacement to identify where the linearized MHD theory begins to break down, and compare that to typical coil tolerances in existing tokamaks. We find that linear theory is valid for existing tolerances, validating the use of perturbative codes to set these tolerances, but with a sufficiently small margin to be of note for future devices with relative tolerances larger than approximately $1\%$ of the minor radius. This study enables engineers to confidently use 3D perturbative models for determining assembly tolerances by providing insight into the correct applications of the theory.

\end{abstract}

%
%
%
\maketitle
%
\ioptwocol

\section{\label{sec:Motivation} Motivation}

Tokamaks are designed to be nominally axisymmetric devices. However, unavoidable non-axisymmetric "error fields" occur due to construction errors, current feeds into the electromagnetic coils, and ferritic materials in the vicinity of the device \cite{luxonAnomaliesAppliedMagnetic2003, wengeAlignmentAssemblyEAST2005, lanctotPathStableLowtorque2016}. Error fields on the order of ten thousand times weaker than the axisymmetric field can reduce plasma rotation through neoclassical torque \cite{ zhuObservationPlasmaToroidalMomentum2006,callenEffects3DMagnetic2011, straitMagneticControlMagnetohydrodynamic2014}, modify the heat flux distribution on the divertor \cite{munarettoAssessmentEquilibriumField2019, munarettoImpactErrorFields2025a}, and create magnetic islands \cite{fitzpatrickInteractionResonantMagnetic1991, lahayeCriticalErrorFields1992} which can reduce confinement and lead to a disruption \cite{henderEffectResonantMagnetic1992, devriesSurveyDisruptionCauses2011,sweeneyMHDStabilityDisruptions2020}. These features make identifying and correcting error fields of the utmost importance in tokamak operation and design, and ultimately set the allowed tolerances for device construction \cite{shibanumaAssemblyStudyJT60SA2013,amoskovAdvancedComputationalModel2018,zumboloOptimalMagnetAssembly2024, pharrErrorFieldPredictability2024}. 

The effects of error fields on the plasma are commonly modeled using 3D perturbed equilibria, which describe how a nominally axisymmetric equilibrium responds to externally applied 3D magnetic fields. This linear formulation can be solved more simply than fully nonlinear equilibria, and several perturbative codes have been developed \cite{liuFeedbackStabilizationNonaxisymmetric2000, liuToroidalSelfconsistentModeling2008, ferraroCalculationsTwofluidLinear2012, parkComputationThreedimensionalTokamak2007a, parkShieldingExternalMagnetic2009, parkImportancePlasmaResponse2009a} and used to predict plasma responses to both error fields resonating in the plasma core and resonant magnetic perturbations in the edge \cite{kimHighestFusionPerformance2024, yangTailoringTokamakError2024}. In this study, we will focus on two main limitations faced when using perturbative models: the choice of reference frame and the validity of the linearized MHD model.

\begin{figure}
    \centering
    \includegraphics[width=\linewidth]{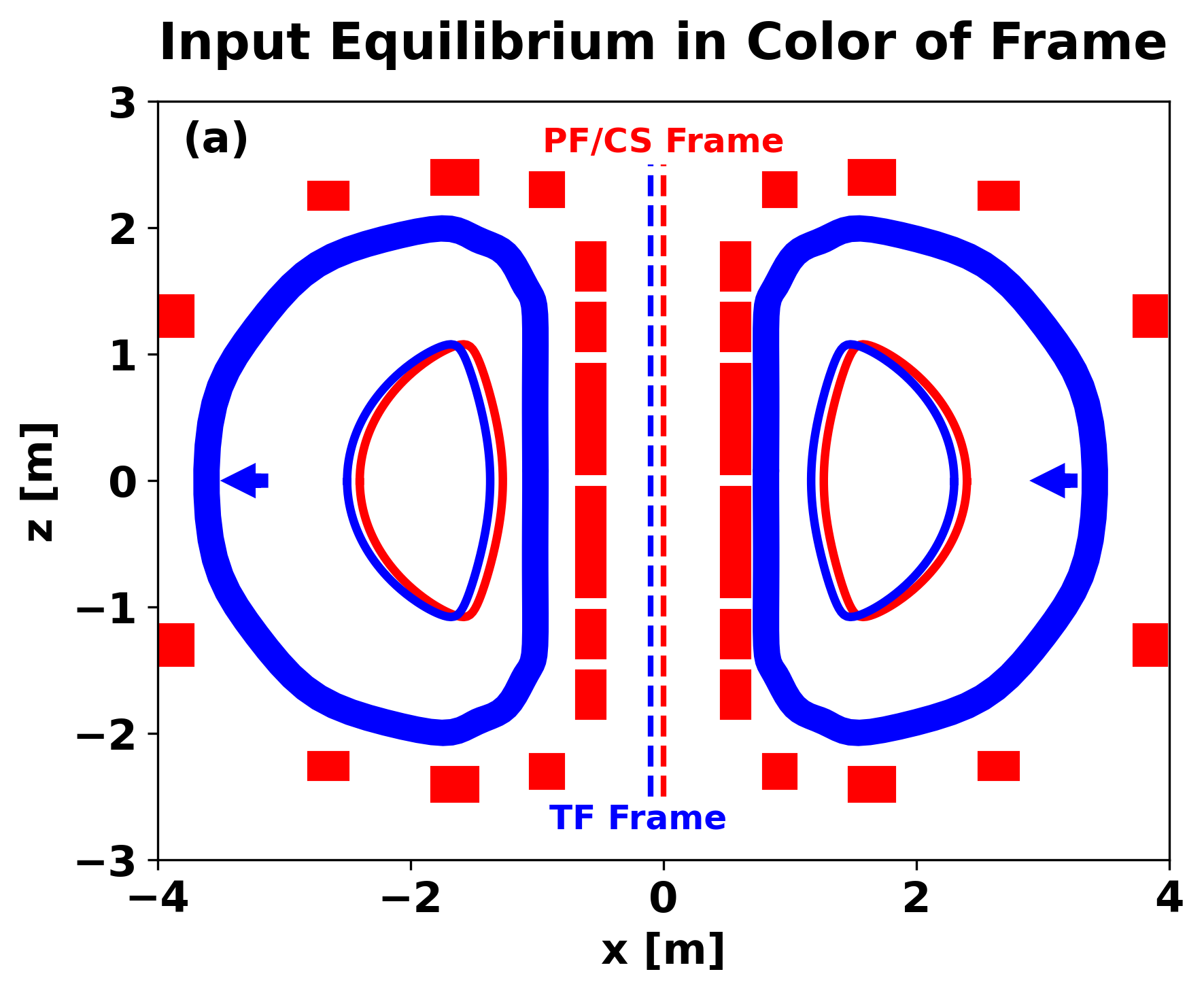}
    \vspace{-0.5em} 
    \includegraphics[width=\linewidth]{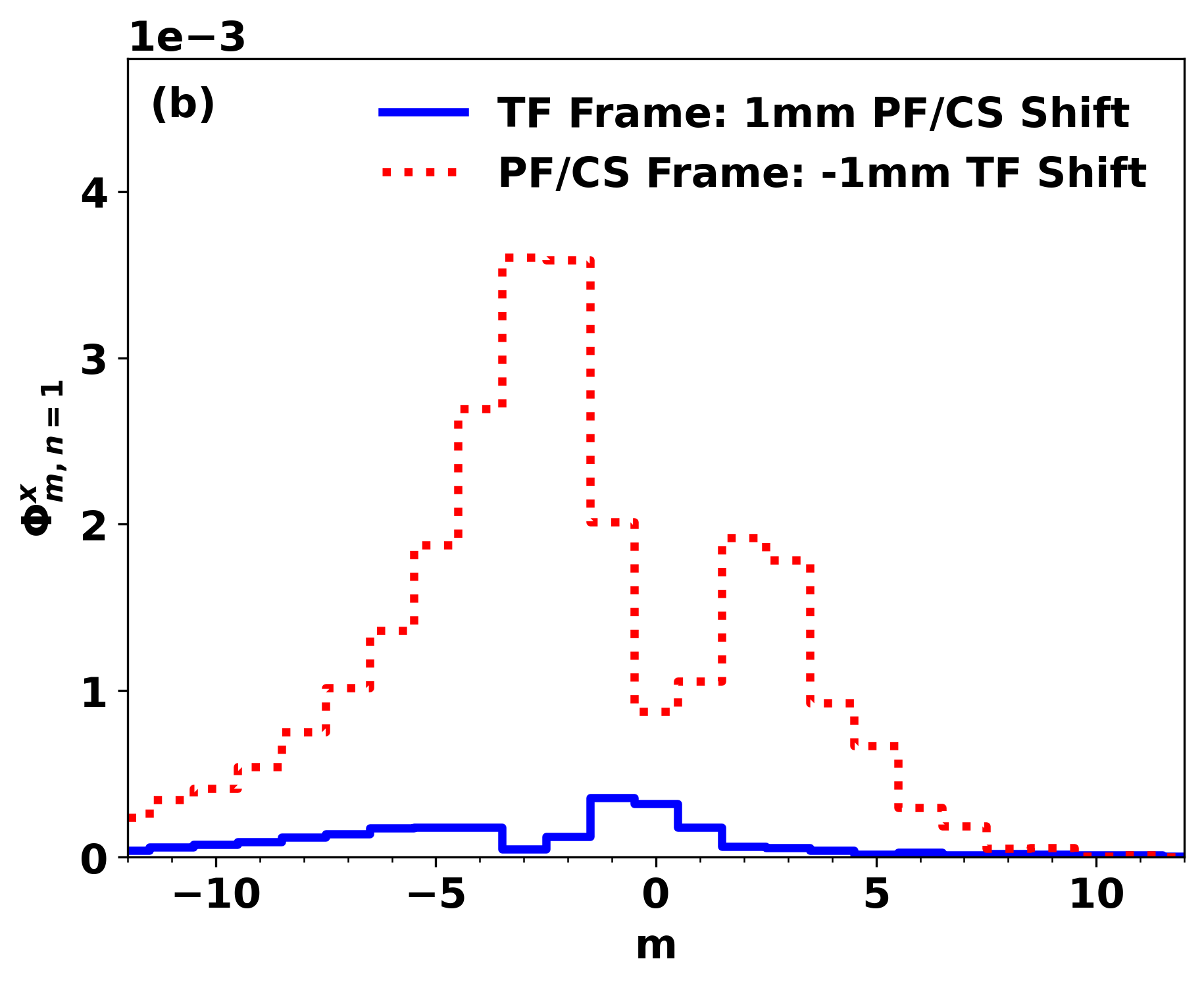}
    \caption{(a) Sample coil set with toroidal field coils in blue rigidly shifted in the $+\hat{x}$ direction. The cross section of the unshifted poloidal field coils and central solenoid is shown in red. (b) Energy normalized applied flux spectrum calculated in GPEC, $\Phi^x_m$, versus poloidal mode number in Hamada coordinates, $m$. The red dashed spectrum is the applied flux on the red surface in (a) from the shifted TF coils, and the blue spectrum is the applied flux on the blue surface from the PF/CS coils.}
    \label{fig:ref_frame}
\end{figure}

\paragraph{Reference Frame} A subtlety that arises when using a perturbed equilibrium code is the correct choice of the reference axis for defining axisymmetry, hereafter referred to as the reference frame. As the equilibrium position is fixed within pertubative codes, this frame is dictated by the input coil positions and varied by rigid shifts of the entire coil set. To illustrate this impact, we consider an example configuration for the SPARC tokamak \cite{creelyOverviewSPARCTokamak2020}, with the cross section in the $x, z$ plane visualized in \cref{fig:ref_frame}(a). In this example, the entire toroidal field (TF) coil set in blue has been rigidly shifted in the $-\hat{x}$ direction during assembly, with the poloidal field (PF) coils and central solenoid (CS) in red installed as specified in the lab frame, the intended axis of symmetry of the device.
When specifying the input 3D coil set, the user assumes that the computed equilibrium will obey the constraints of the linearized theory, which has a singular point at the magnetic axis and forces the axis displacements to go to zero. 
The perturbed equilibrium must therefore retain the same magnetic axis as the initial axisymmetric equilibrium, requiring the user to specify the coils in the frame that preserves the magnetic axis. In other words, the shifted coil set should be specified relative to the true magnetic axis location. 

In our example, the user must assume if the plasma axis is fixed with respect to the PF/CS coils, TF coils, or somewhere in between. For the first case, which we denote the PF/CS frame, we assume that the plasma retains the original axis of symmetry (red dashed line in \cref{fig:ref_frame}(a)), and the input equilibrium in red is perturbed due to the shifted TF coils. As the PF/CS coils are unshifted, the PF/CS frame is equivalent to the lab frame in this case. In the TF frame, the user assumes that the plasma shifts with the TF coils, resulting in an input equilibrium that is axisymmetric with respect to the TF centroid (blue dashed line in \cref{fig:ref_frame}(a)) with the same plasma boundary as the original equilibrium; the 3D coil set is then rigidly shifted relative to the plasma, and the nonaxisymmetric field becomes the PF/CS coils shifted in the $+\hat{x}$ direction. For all frames in between, the correct choice of input coils would include both the TF and PF/CS coils shifted in opposite directions, with a relative shift between them equal to the original TF shift. This assumption of which reference frame is correct will result in differing applied 3D field spectra, as shown in \cref{fig:ref_frame}(b) where we plot the normalized applied flux on the plasma surface, 
\begin{equation}
    \Phi^x_{mn} = \frac{1}{(2\pi)^2} \iint e^{i(n\phi - m\theta)} \frac{\delta \vec{B} \cdot \hat{n}}{\vec{B} \cdot \nabla \theta} d\theta d\phi,
\end{equation}
in both the TF frame and PF/CS frame. Here, $\vec{B}$ is the magnetic field, $\phi/\theta$ the toroidal/poloidal angles, $n/m$ the toroidal/poloidal mode numbers, and $\hat{n}$ the surface unit normal. The applied field spectrum determines the plasma response \cite{parkImportancePlasmaResponse2009a}, indicating that the response differs in these two frames despite being the same physical coil set translated in space. The magnitude of the response difference will depend on the overlap of the applied spectra with the dominant mode of the plasma, which is the externally applied field spectrum that most strongly drives a resonant response \cite{parkControlAsymmetricMagnetic2007, loganEmpiricalScalingError2020}. This difference can have effects on various aspects on device operation, such as the optimal currents in error field correction coils determined from perturbative models. The choice of reference frame has previously been assumed, with little work performed to validate the assumption. This work seeks to clarify this assumption and provide guidance to future studies.

\paragraph{Linearity}
Even if the correct reference frame is assumed and the magnetic axis criterion is met, perturbative codes are still limited by the underlying model; at some magnitude of asymmetry in the coil set, the linearized theory is no longer valid, as it assumes the displacements from the initial equilibrium are small compared to other equilibrium parameters. While extensive studies have been performed in comparing perturbed with fully 3D equilibria \cite{wingenConnectionPlasmaResponse2015, turnbullComparisonsLinearNonlinear2013, reimanTokamakPlasmaHigh2015, kingExperimentalTestsLinear2015}, little work has been performed in analyzing when the two no longer agree.

\paragraph{}
In this paper, we employ fully nonlinear 3D equilibria to address the above limitations inherent to perturbative equilibrium models, specifically for the case where subsets of the coil configuration become misaligned during assembly. Several nonlinear equilibrium solvers exist \cite{hirshmanSteepestdescentMomentMethod1983, reimanCalculationThreedimensionalMHD1986, suzukiDevelopmentApplicationHINT22006, hudsonComputationMultiregionRelaxed2012,dudtDESCStellaratorEquilibrium2020}, and are most often used within the stellarator field. Without the linearized MHD assumptions, these models can identify the nonlinear 3D equilibrium state provided a vacuum magnetic field produced from coils and plasma profiles describing the current and pressure. Most notably for this study, we identify the magnetic axis location from nonlinear equilibria in various coil set geometries, which can provide insight into the correct choice of reference frame to use in perturbative codes. We will consider primarily $n=1$ error fields, which have been shown to be harmful to confinement \cite{butteryErrorFieldExperiments2000, straitMeasurementTokamakError2014, lanctotImpactToroidalPoloidal2017} and are unique in that they require the choice of the correct reference frame, as there is no equivalent reference frame for $n \geq 2$. In \cref{sec:Codes}, we describe the specific perturbative and nonlinear equilibrium solver used in this work (GPEC and VMEC, respectively). In \cref{sec:RadialRefFrames}, we identify the correct reference frame for use in perturbed equilibrium codes with $n=1$ radial coil shifts, specifically for the case of the SPARC tokamak with offset PF/CS and TF coil sets as well as in NSTX-U with a shifted/tilted centerpost containing the CS coils and inner legs of the TF coils. In \cref{sec:Sensitivity}, we demonstrate the sensitivity of perturbative models to this choice of reference frame for the NSTX-U example by calculating the island width in both the lab frame and correct frame.  In \cref{sec:GeneralRefFrames}, we analyze $n=1$ vertical shifts of the TF coils and discuss the implications for the choice of reference frame for more general coil asymmetries. In \cref{sec:Linearity}, we use displacements computed from nonlinear equilibria to identify the breakdown of perturbative theory in the correct reference frame as a function of the magnitude of the coil set shifts.

\section{\label{sec:Codes} Description of Equilibrium Solvers}

For our perturbative model, we use the Generalized Perturbed Equilibrium Code (GPEC) \cite{parkComputationThreedimensionalTokamak2007a, parkShieldingExternalMagnetic2009, parkImportancePlasmaResponse2009a}, which is built off of the ideal MHD stability code DCON \cite{glasserDirectCriterionNewcomb2016b}. DCON solves for the displacements $\vec{\xi}$ which minimize the perturbed plasma potential energy, 
\begin{equation}
    \delta W = \frac{1}{2} \int_V \vec{\xi} \cdot F\left(\vec{\xi}\right) dV,
    \label{eq:deltaW2}
\end{equation}
where $F(\vec{\xi})$ is the linearized ideal MHD force operator, 
\begin{equation}
    F\left(\vec{\xi}\right) = \delta \vec{J} \times \vec{B}_0 + \vec{J}_0 \times \delta \vec{B} +  \nabla \delta p.
\end{equation}
These displacements describe the set of possible perturbed equilibria which satisfy force balance. GPEC uses linear combinations of these force-free eigenmodes from DCON to form the actual perturbed equilibrium from a forcing term of external 3D vacuum fields. 

GPEC also employs a densely packed radial grid around rational surfaces, allowing the determination of response metrics at rational flux surfaces. For example, one of the main quantities of interest in linear codes is
\begin{equation}
    \Delta'_{mn} = \frac{\partial}{\partial s} \left(\frac{\delta \vec{B} \cdot \hat{n} |\nabla s|}{\vec{B} \cdot \nabla \phi} \right)_{mn},
    \label{eq:Deltaprime}
\end{equation}
which parametrizes the $m,n$ Fourier harmonic of the parallel  current that forms to screen out the externally applied magnetic perturbations on each rational surface \cite{fitzpatrickInteractionResonantMagnetic1991,boozerPerturbedPlasmaEquilibria2006, nuhrenbergMagneticSurfaceQualityNonaxisymmetric2009, loganEmpiricalScalingError2020}.  Here, $s$ is a radial-like coordinate. This metric is directly used in linear tearing theory to quantify the free energy that can drive magnetic island formation.

For our nonlinear model, we use the Variational Moments Equilibrium Code (VMEC) \cite{hirshmanSteepestdescentMomentMethod1983}, commonly used in stellarator optimization and design. Unlike the perturbed ideal MHD potential energy solved in DCON, VMEC minimizes the total ideal MHD energy without assumptions on the size of the 3D fields.
As VMEC also includes only ideal effects when computing equilibria, we can directly compare the outputs from GPEC and VMEC without considering differences due to kinetic or resistive physics. 
VMEC does not fully resolve the plasma response at rational surfaces \cite{lazersonVerificationIdealMagnetohydrodynamic2016}, so we will not directly compare resonant metrics such as \cref{eq:Deltaprime} between VMEC and GPEC. Instead, we primarily use VMEC to identify the magnetic axis location for various coil set geometries, which can provide insight into the correct choice of reference frame to use in perturbative codes such as GPEC to obtain an accurate prediction of \cref{eq:Deltaprime}. We also evaluate the normal displacement of the magnetic field within VMEC to determine the transition from linear to nonlinear equilibria, which has been shown to converge at sufficiently high radial resolution \cite{wingenConnectionPlasmaResponse2015}.

\section{\label{sec:RadialRefFrames} Effect of Radial Shifts on the Reference Frame}

We begin with the case of determining the correct reference frame for $n=1$ radial coil set shifts. We will consider two different examples based on different mechanical assembly strategies: the SPARC and NSTX-U tokamaks. We chose to use both equilibria to investigate how the correct reference frame depends on the aspect ratio and the coil geometry, i.e. if the TF coil set is treated as rigidly connected or separated into inner and outer legs. SPARC's TF coils are not demountable and built before device assembly, so we treat them as independent to the PF and CS coils. In NSTX-U, the centerpost is instead built separately and contains the CS coils with the center rods of the TF coils inside, so we treat the centerpost as independent of the PF coils and demountable outer legs of the TF coils. We run VMEC in free boundary mode with coil sets generated in the Simsopt framework \cite{landremanSIMSOPTFlexibleFramework2021}. The specifics regarding how the coil sets were generated are described in \cref{app:coils}.

\subsection{SPARC: Independent TF and PF/CS Coils}

\begin{figure}
    \centering{}
    \includegraphics[width=1.0\linewidth]{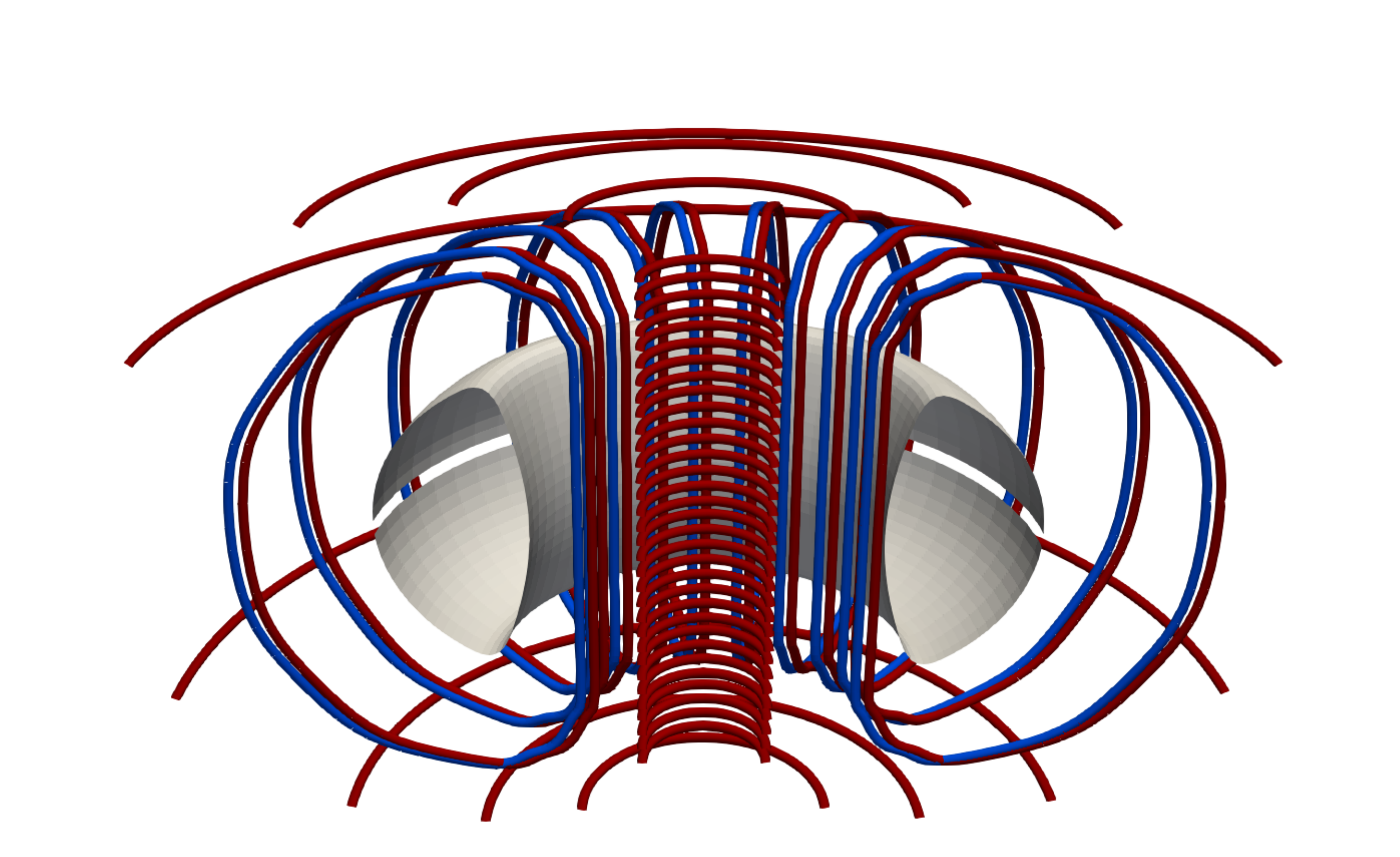}
    \caption{Half-view of the SPARC coil set with a $-100~\mathrm{mm}$ radial toroidal field coil set shift. The shifted coils are in blue and unshifted coils in red. We remove some filaments used to represent the CS and TF coils for clarity, and plot the unshifted equilibrium surface in grey. }
    \label{fig:sparccoils}
\end{figure}

We analyze a SPARC L-mode equilibrium from the time-dependent evolution that leads to the primary reference discharge \cite{rodriguez-fernandezOverviewSPARCPhysics2022, bodySPARCPrimaryReference2023}, as the low density makes the plasma more susceptible to error field penetration \cite{loganRobustnessTokamakError2020}. We assume independent shifts of the PF/CS and TF coil sets (as described in the introductory example). \Cref{fig:sparccoils} shows an example 3D coil set for SPARC with TF coils shifted $-100~ \mathrm{mm}$.

\begin{figure}
    \centering{}
    \includegraphics[width=1.0\linewidth]{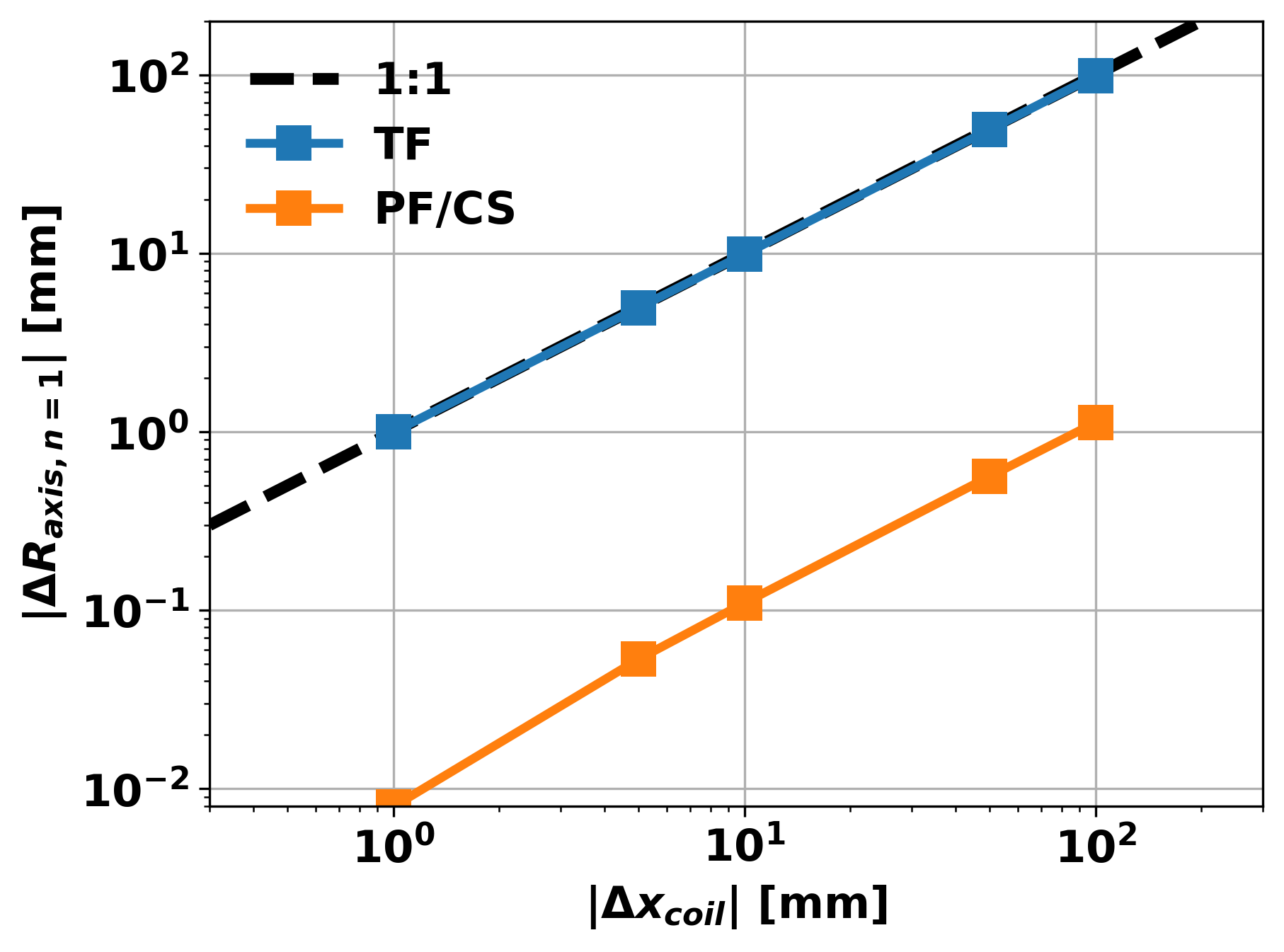}
    \caption{$n=1$ component of the magnetic axis calculated by VMEC as a function of shift for both the TF (blue) and PF/CS (orange) coils sets. The 1:1 line (dashed black) represents the equilibrium axis being fixed to the coil set centroid.\label{fig:n1_TF_vs_PFCS}}
\end{figure}

\begin{figure}
    \centering{}
    \includegraphics[width=1.0\linewidth]{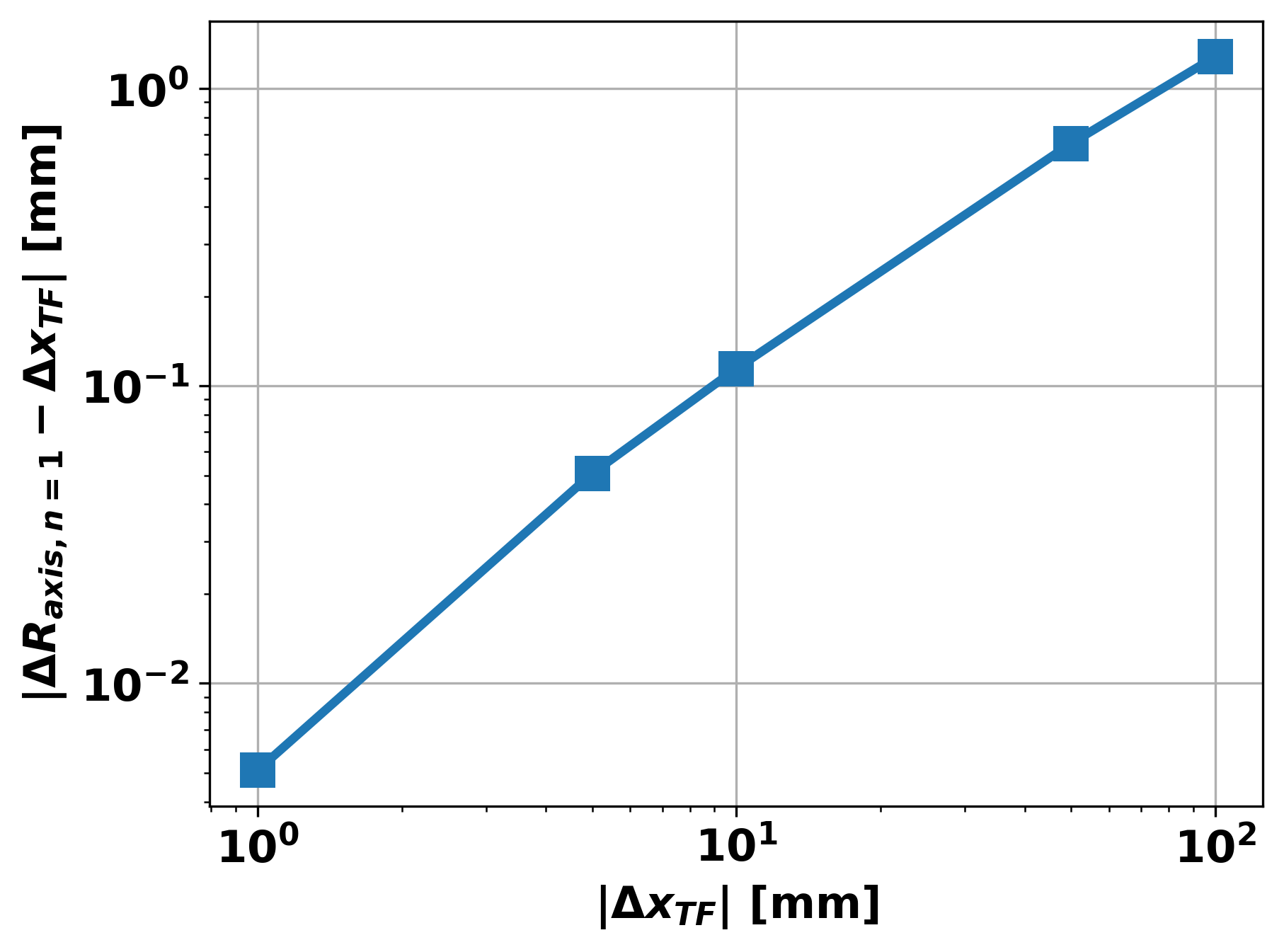}
    \caption{Difference between the $n=1$ equilibrium axis from VMEC and TF shift as a function of the TF shift. }
    \label{fig:n1_diff}
\end{figure}

For the SPARC equilibrium, we scan the radial shift of both the TF and PF/CS coil sets for shifts from $1$ to $100 ~\mathrm{mm}$, calculating a new VMEC equilibrium for each coil set and plotting the resulting $n=1$ component of the magnetic axis major radius in \cref{fig:n1_TF_vs_PFCS}. As a reminder, a TF shift corresponds to assuming the PF/CS frame and a PF/CS shift corresponds to assuming the TF frame when running a perturbative model. We also plot the 1:1 line, representing the equilibrium axis being fixed with respect to shifts of the coil set. This analysis shows that the equilibrium shift can be approximated by the shift of the TF coil set, which nearly overlaps the 1:1 line. The axis variations from the TFs are also two orders of magnitude larger than the PF and CS coil set for the entire range of shifts plotted. The correct reference frame can thus be best determined by the centroid of the TF coil set, and perturbative models should perturb about an equilibrium that is axisymmetric in this frame. We show the discrepancy due to this approximation in \cref{fig:n1_diff}. As expected from \cref{fig:n1_TF_vs_PFCS}, the difference in the magnetic axis of the shifted equilibrium and the centroid of the TF coil set is two orders of magnitude less than the shift, and can be attributed to the shift caused by the PF and CS coil set. This difference directly sets the error in assuming the TF reference frame as opposed to the "true" frame, which is set by the 3D magnetic axis and must be determined from the nonlinear equilibrium.

Based on \cref{fig:ref_frame}(b), this result seems intuitive; the TF coils produced much larger applied fluxes than the PF/CS coils for the same shift, so the plasma responds more strongly to the TF error field and shifts in that direction. However, \cref{fig:ref_frame}(b) indicates an order of magnitude difference in the applied flux, while the use of nonlinear equilibria demonstrates more than two orders of magnitude difference in the effect on the magnetic axis. We will also show in \cref{sec:GeneralRefFrames} that the amplitude alone does not solely indicate the movement of the axis, further motivating the use of nonlinear models to address these questions. 

In order to assess the relative error from assuming the TF frame in perturbative models, we run GPEC in three reference frames for a $1~\mathrm{mm}$ relative shift: the PF/CS, TF, and true frame. The true frame in this case is shifted by $0.995~\mathrm{mm}$ in the TF shift direction based on the results of \cref{fig:n1_TF_vs_PFCS}; the input coils to GPEC were then the TF coils offset by $-0.005~\mathrm{mm}$ and PF/CS coils by $0.995~\mathrm{mm}$, both in the $\hat{x}$ direction. In the true frame, the axis remains unshifted due to the applied fields, retaining full validity of the perturbed equilibrium model. 

\begin{figure}
    \centering{}
    \includegraphics[width=1.0\linewidth]{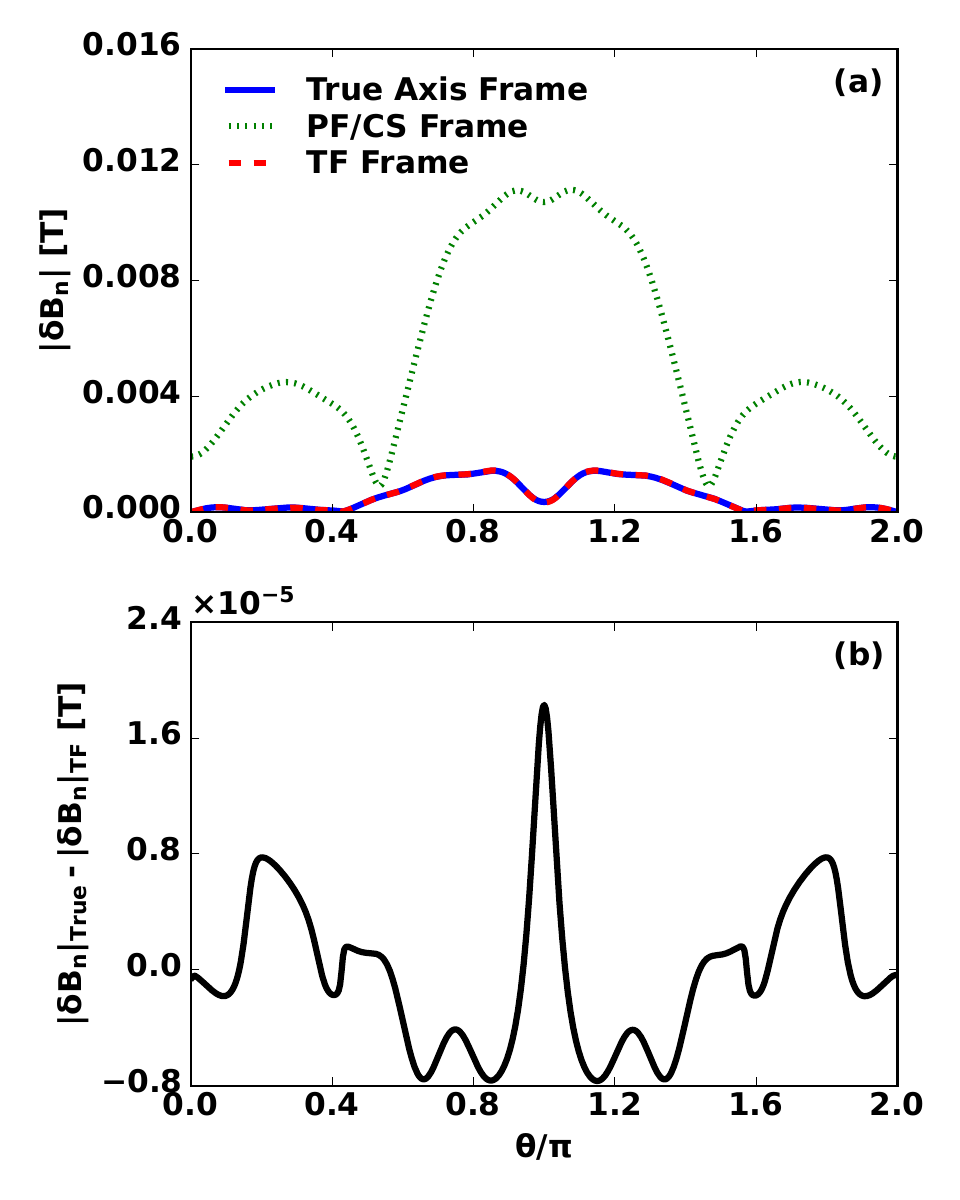}
    \caption{(a) Perturbed normal field at the last-closed flux surface calculated by GPEC for a $1~\mathrm{mm}$ PF/CS shift (TF frame) in dotted green, $-1~\mathrm{mm}$ TF shift (PF/CS frame) in dashed red, and $-0.005~\mathrm{mm}$ TF shift and $0.995~\mathrm{mm}$ PF/CS shift (true frame) in solid blue. (b) Difference between the normal field computed in the true frame and approximate TF frame.}
    \label{fig:corrected_frame}
\end{figure}

We plot the amplitude of the perturbed normal field at the last closed flux surface, $\delta B_n$ for each of the three frames in \cref{fig:corrected_frame}(a). It is clear that the PF/CS frame produces a very different normal field, as expected from our analysis of \cref{fig:n1_TF_vs_PFCS}. This result is unphysical, as we have shown this normal field would contribute to the magnetic axis shift rather than a direct error field, invalidating the perturbative model. However, the TF and true frame closely agree, and we plot the difference between the two in \cref{fig:corrected_frame}(b). This error can be interpreted as the error due to assuming the TF reference frame, as opposed to the true frame that must be computed from the nonlinear equilibrium. The maximum error is around $1.8\times10^{-5}~\mathrm{T}$, whereas the maximum normal fields in the correct frame are on the order of $10^{-3}~\mathrm{T}$, indicating that assuming the TF frame only introduces small errors into perturbative calculations. 

Our findings show that the TF frame is a good approximation for the true reference frame when running perturbative codes. As a result, the radial TF $n=1$ asymmetries that normally produce large shifts of the magnetic axis and invalid the perturbative model can be translated into a new frame which preserves the magnetic axis location, without further computation of nonlinear equilibria. In this TF frame, a shift of the TF coils in the lab frame is modeled in perturbative codes by an opposite shift of the PF/CS coil set.

\subsection{NSTX-U: Independent Centerpost and Outer TF Legs/PF Coils}

\begin{figure}
    \centering{}
    \includegraphics[width=1.0\linewidth]{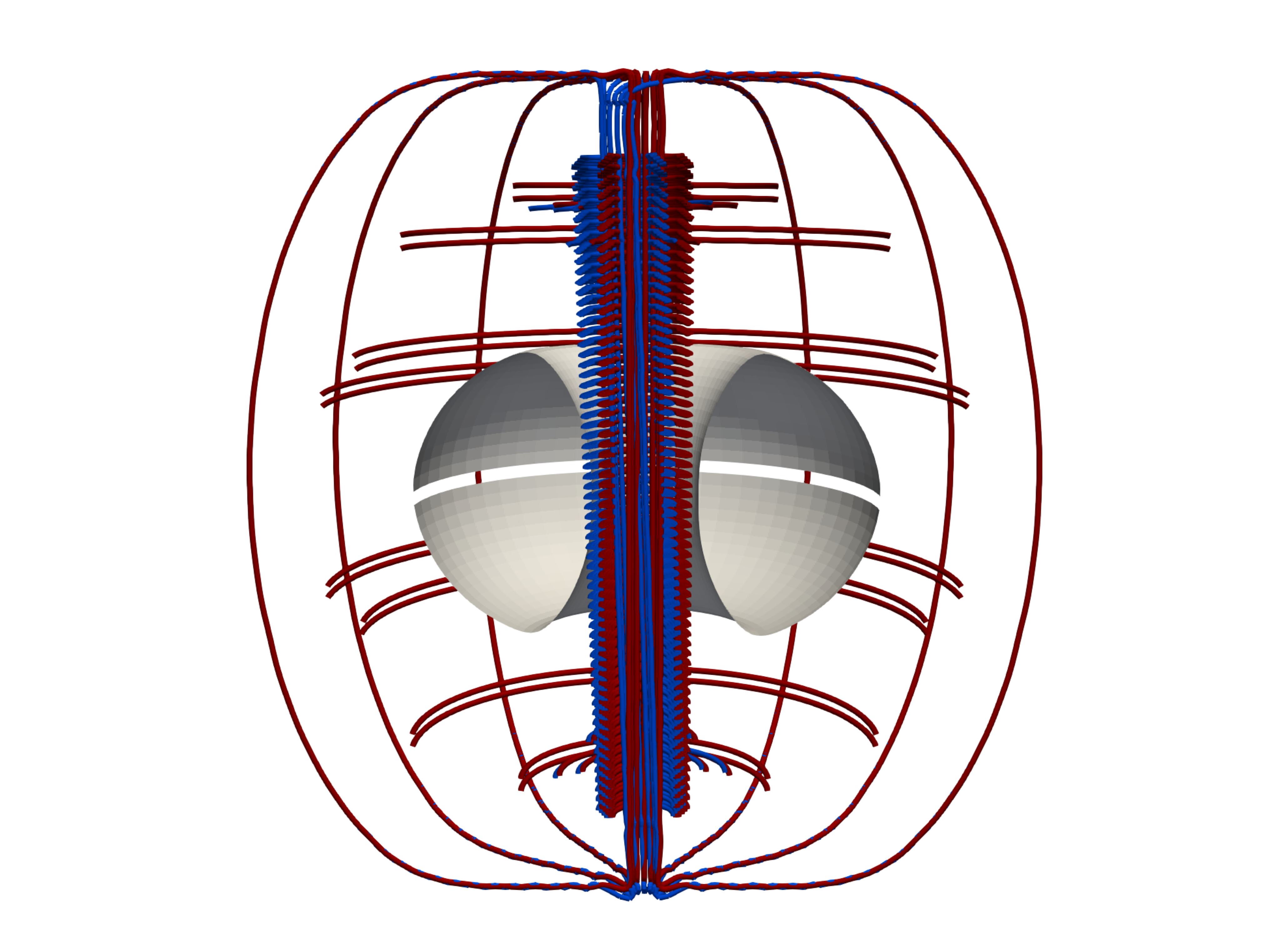}
    \caption{Cross section of the shifted (blue) and unshifted (red) coil sets for the NSTX-U configuration with a $2^\circ$ tilt of the centerpost. We plot the unshifted equilibrium surface in grey. }
    \label{fig:nstxucoils}
\end{figure}

We now consider the case of an independent centerpost and outer TF legs/PF coils in NSTX-U \cite{berkeryNSTXUResearchAdvancing2024}, to observe the effect of separating the contribution of the inner and outer legs of the TF, as well as if aspect ratio affects our conclusions. The centerpost consists of the inner legs of the TF coils as well as the CS. We analyze an up-down symmetric NSTX-U representative example equilibrium with both rigid shifts and tilts of the centerpost. For the tilts, the pivot point is set to the bottom of the device \cite{menardOverviewNSTXUpgrade2017, battagliaScenarioDevelopmentCommissioning2018,ferraroErrorFieldImpact2019}. We show an example 3D coil set for a $2^\circ$ centerpost tilt in \cref{fig:nstxucoils}.

\begin{figure}
    \centering{}
    \includegraphics[width=1.0\linewidth]{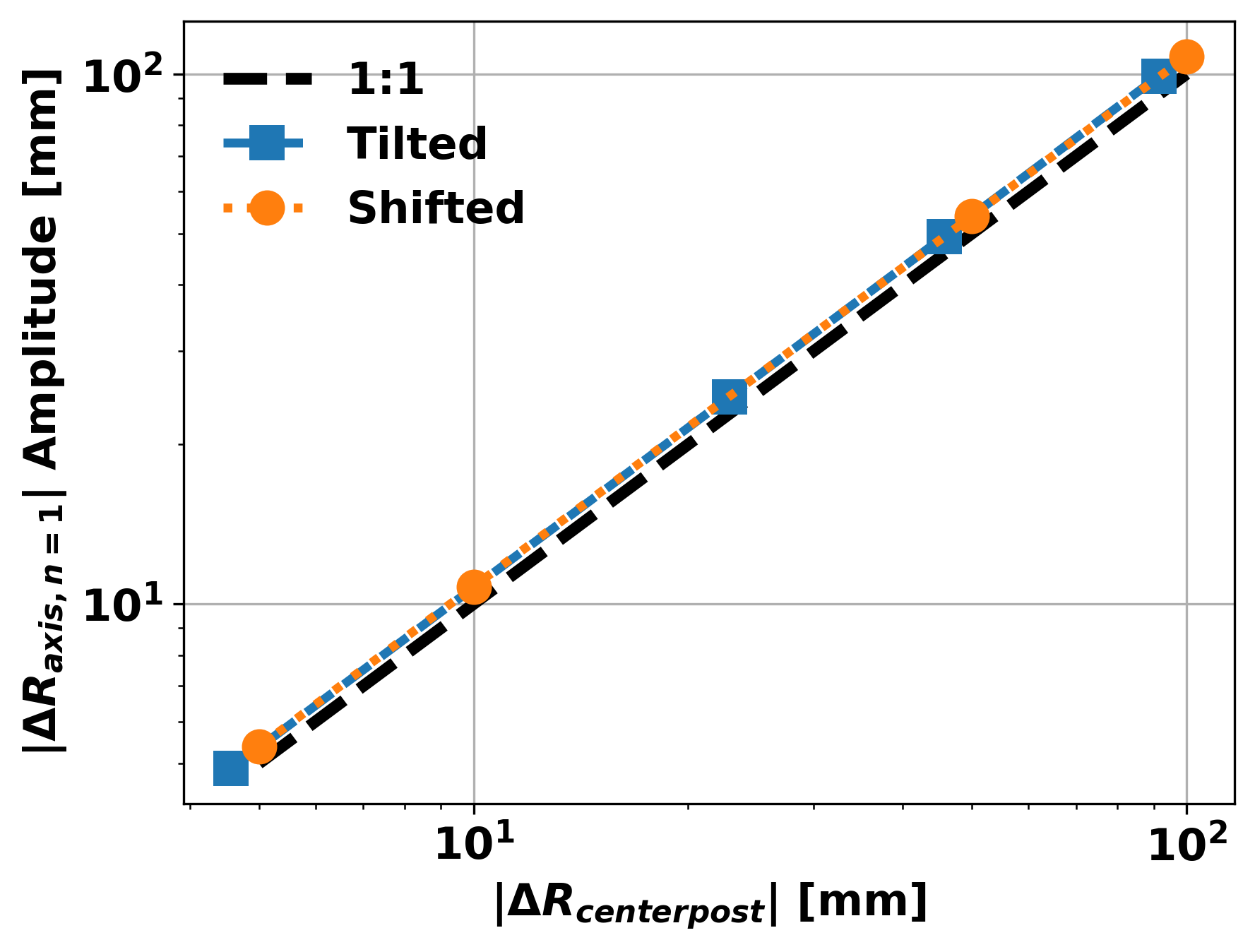}
    \caption{Variation of the magnetic axis determined from VMEC as a function of the radial variation of the centerpost for both centerpost shifts and tilts (pivoting about the bottom of the device). The tilts are plotted as an equivalent radial shift of the $Z=0$ (midplane) section of the centerpost.  \label{fig:n1_shift_v_tilt_nstxu}}
\end{figure}

We perform the same analysis as the SPARC equilibrium, plotting the $n=1$ radial component of the magnetic axis determined from fully 3D VMEC equilibria as a function of both types of coil set asymmetries. The results are plotted in \cref{fig:n1_shift_v_tilt_nstxu}, where we have plotted the tilts in the form of an equivalent shift of the centerpost midplane, 
\begin{equation}
    |\Delta R_{\mathrm{centerpost}}(Z=0)| = Z_{\mathrm{max}} \tan(\theta) \approx Z_{\mathrm{max}} \theta,
\end{equation}
with $Z_{max}$ is the distance between the midplane and the pivot point (half the total height of the device) and $\theta$ is the tilt angle. We can see that the equilibrium is nearly fixed with the radial location of the centerpost midplane for both cases. For the shift, this implies that the inner TF legs are the dominant factor when setting the reference frame, as opposed to the outer legs, or else the normal field from the outer legs would have placed the magnetic axis somewhere below the 1:1 line. For the tilt, this signifies that a tilt pivoting about the bottom of the device can be decomposed into a rigid $n=1$ radial shift of the equilibrium, which modifies the reference frame, and a tilt about the midplane of the device, which produces the error field in the true frame. Note that this is analogous to the results from the SPARC equilibrium study, where the TF coils shifted the equilibrium, and the PF/CS coils, now shifted relative to the plasma in the opposite direction, create the error field. The tilted centerpost is just more complex in that the centerpost shift introduces both a rigid shift component and an error field component, so that the final error field is determined by both the outer TF legs/PF coils shifted relative to the new equilibrium and the tilted centerpost in the correct frame. The determination of the reference frame thus remains agnostic to the assembly strategy and aspect ratio of the device; we will show in \cref{sec:GeneralRefFrames} that the reference frame is primarily determined by the spectrum of the applied field, which varies with the source of the misalignment. 

\section{\label{sec:Sensitivity} Linear Model Sensitivity to the Reference Frame}

To illustrate the importance of including this effect in perturbative models, we run GPEC with two NSTX-U coil sets which would result in the same error fields in a non-perturbative code: a centerpost tilted by $1^\circ$ in the lab frame, and the same case but in the TF radial frame. The coil set in the radial frame is obtained by rigidly shifting the entire coil set in the lab frame by $Z_{\mathrm{max}}\tan(\theta)$ in the direction opposite to the tilt to place the midplane of the centerpost at $R=0$; we know that the new magnetic axis with this shifted coil set will remain approximately axisymmetric based on our findings in \cref{fig:n1_shift_v_tilt_nstxu}. Note that this shift to correct the reference frame transforms the centerpost tilt about the bottom of the device into a tilt about the midplane, with additional 3D fields introduced from the newly shifted outer TF legs and PF coils. 

\begin{figure}
    \centering{}
    \includegraphics[width=1.0\linewidth]{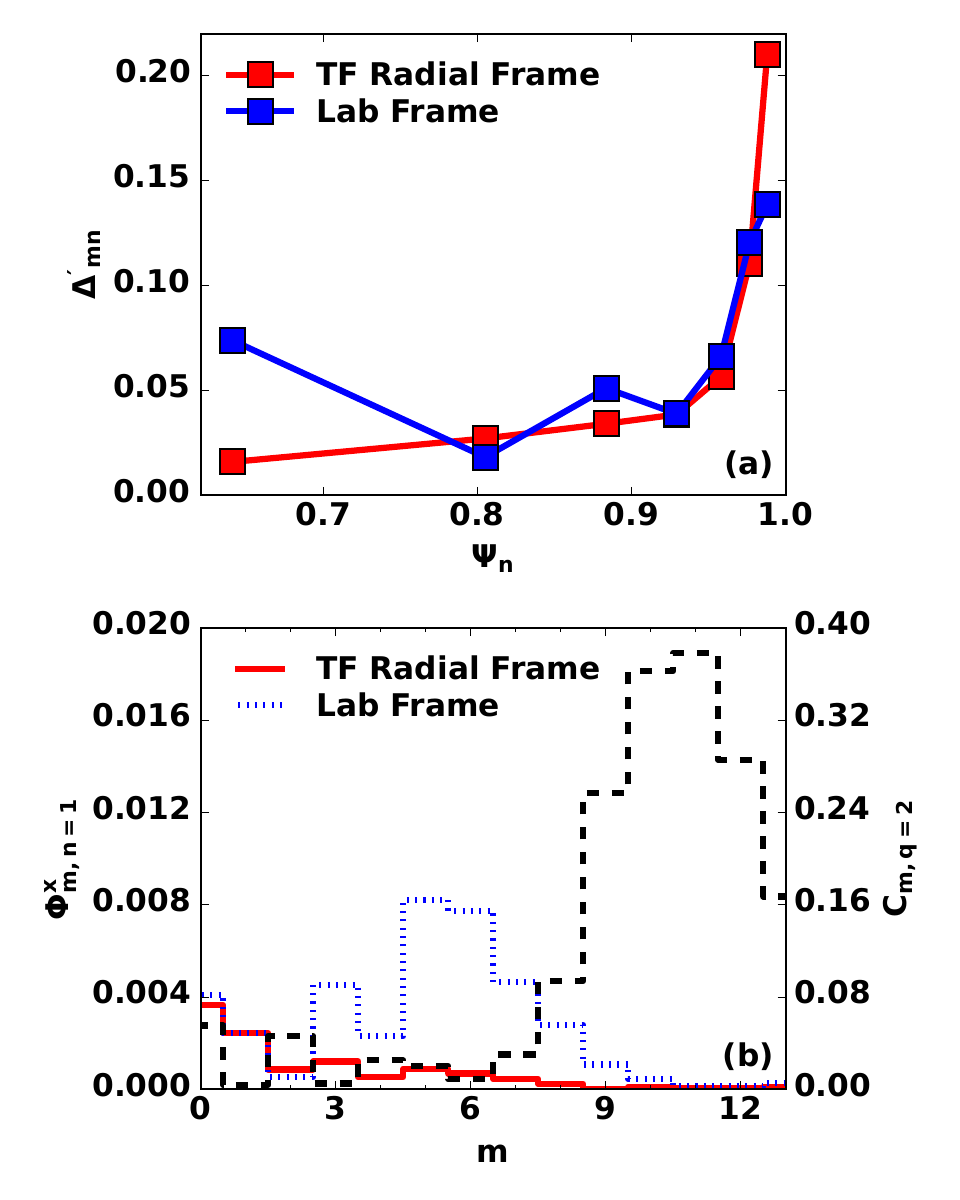}
    \caption{(a) Linear prediction of $\Delta'$ (\cref{eq:Deltaprime}) from GPEC with $n=1$ at several rational surfaces for a $1^{\circ}$ tilt of the centerpost. The blue case is in the lab frame, while the red case is in the TF radial (correct) frame. (b) Energy normalized applied flux spectrum for each frame (left axis) along with the coupling spectrum between the applied flux and flux on the $q=2$ surface (right axis)}
    \label{fig:nstx_deltaprime}
\end{figure}

As discussed in the context of \cref{fig:ref_frame}(b), correcting the reference frame modifies the applied flux on the plasma, which can modify the plasma response depending on how this change overlaps with the applied modes to which the plasma is most sensitive. To observe the effect for this example, we plot the value of $\Delta'$ predicted by GPEC according to \cref{eq:Deltaprime} at several rational surfaces in \cref{fig:nstx_deltaprime}(a), starting with $q=2$. It is clear that these two GPEC runs predict a vastly different plasma response, with $\Delta'$ differing by a factor of around $4$ at the $q=2$ surface, which is typically the most important for determining error field impact on plasma confinement. The discrepancy is due to differences in the applied flux in the lab and radial frame in spectral regions where the coupling between the applied flux and flux on the resonant surfaces is large; we plot a portion of the applied flux spectra for each frame along with the coupling spectrum at the $q=2$ surface, $C_{m, q=2}$, in \cref{fig:nstx_deltaprime}(b) to illustrate this effect. The region where both the coupling spectrum and the differences and applied fluxes are the largest is for poloidal mode numbers between approximately $m=6$ to $m=9$. This example emphasizes the importance of choosing the correct reference frame in perturbative models to obtain the correct applied field spectrum and corresponding plasma response. 

\section{\label{sec:GeneralRefFrames} Effect of General Applied Fields on the Reference Frame}

Up to this point, we have demonstrated that radial shifts of the TF coils, specifically the inner legs, lead directly to shifts of the equilibrium and do not introduce error fields directly, with the error fields instead coming from the shift of the remaining coils relative to the shifted equilibrium. This leads to the question: do all shifts of the TF coil have this same effect, or are there specific asymmetries which do not have a strong impact on the equilibrium axis? 

\begin{figure}
    \centering{}\includegraphics[width=1.0\linewidth]{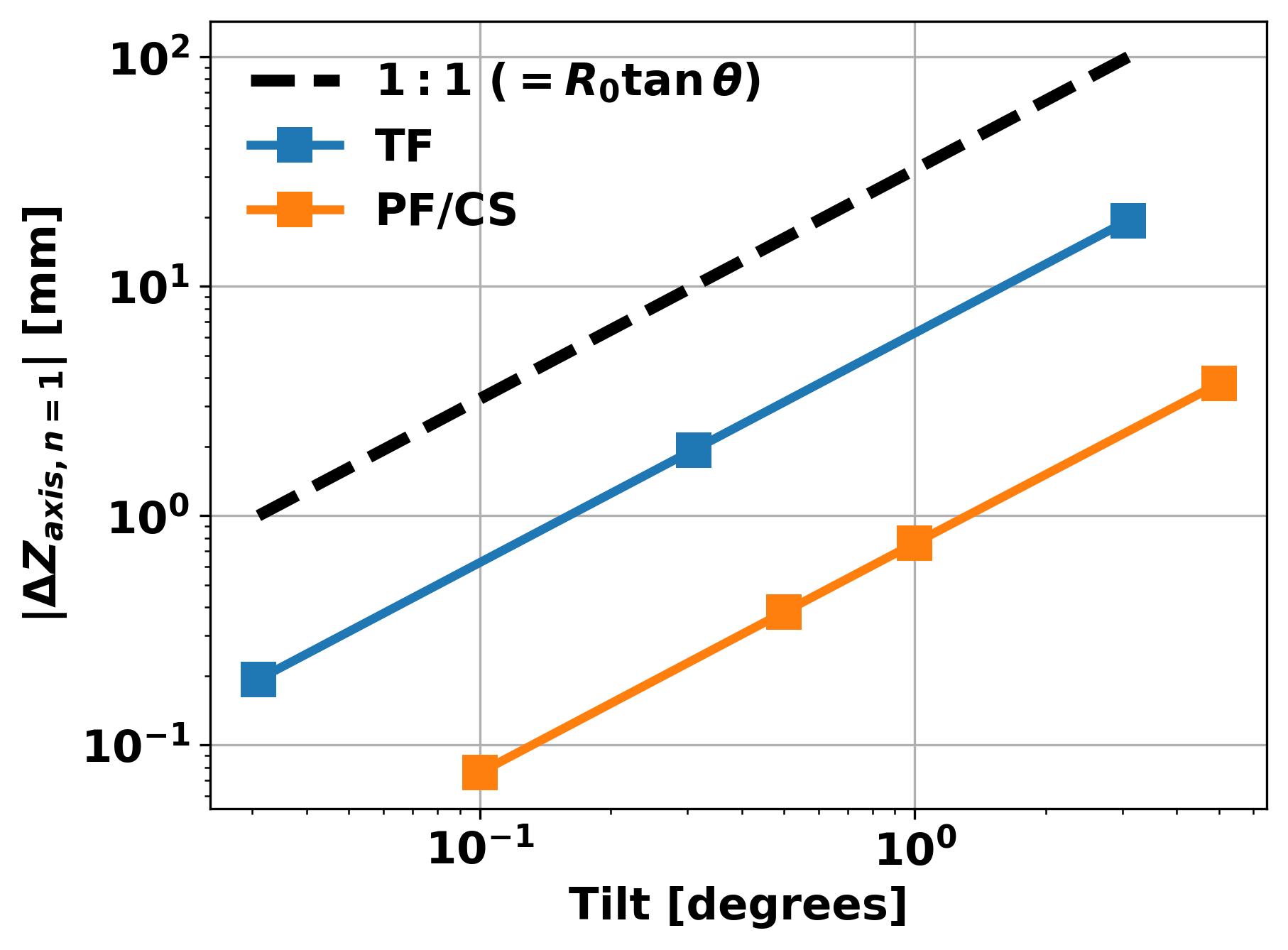}
    \caption{Variation of the $n=1$ vertical component of the magnetic axis calculated from VMEC equilibria for both vertical shifts in the TF centroid and tilts of the PF/CS coils. We convert the TF shifts to an effective tilt with respect to the equilibrium major radius. The 1:1 line is shown, representing the plasma being fixed to the coil tilt. 
    \label{fig:n1_vertical}}
\end{figure}

To address this question, we consider a $n=1$ vertical shift of the TF coil set where the $Z$ centroid, $Z_0$, of each TF coil is modified by
\begin{equation}
    Z_0 = \Delta Z \cos(\phi),
\end{equation}
where $\Delta Z$ is the amplitude of the shift and $\phi$ is the toroidal angle. This gives an effective tilt of the TF coil set with respect to the equilibrium as
\begin{equation}
    \theta = \arctan(\Delta Z / R_0),
\end{equation}
with $R_0$ the equilibrium major radius. We directly compare VMEC equilibria from the shifted TF coils to tilted PF/CS coils sets for the SPARC equilibrium in \cref{fig:n1_vertical}. We again plot the 1:1 line, representing the case where the equilibrium $n = 1 ~Z$ component of the magnetic axis is fixed with the respective coil set. 

Comparing to \cref{fig:n1_TF_vs_PFCS}, it is clear that the radial and vertical TF coil shifts do not produce the same effect, with the equilibrium axis located around halfway between the unperturbed axis and the new TF centroid. The PF/CS coil sets produce axis displacements two orders of magnitude less than a 1:1 shift similar to the radial case. These results indicate that there is no clear choice of reference frame for vertical shifts of the coil sets that can be assumed without knowledge of the nonlinear equilibrium characteristics. Without a reference frame correction, the validity of perturbative models will be limited to regimes where the axis displacement due to the vertical shift is a small parameter. 


These findings indicate a more general trend in the characteristics of normal fields that modify the reference frame and those which induce error fields. For example, in \cref{fig:ref_frame}(a) we saw that the radial TF shift produces normal field amplitudes $10X$ larger than the radial PF/CS shift. This difference in magnitude indicates why the TF frame was correct for the radial shift case. If we transform this spectrum into real space, we also find that it is nearly twice as large on the inboard side than the outboard, giving intuition on the NSTX-U findings by explaining why shifts of the TF inner legs are more impactful than the outer legs for determining the frame. However, this magnitude does not tell the whole story, as the TF and PF/CS vertical shifts produce comparable maxima in the applied fields in real space, but \cref{fig:n1_vertical} showed that the axis displacement differs by an order of magnitude.

\begin{figure}
    \centering{}
    \includegraphics[width=1.0\linewidth]{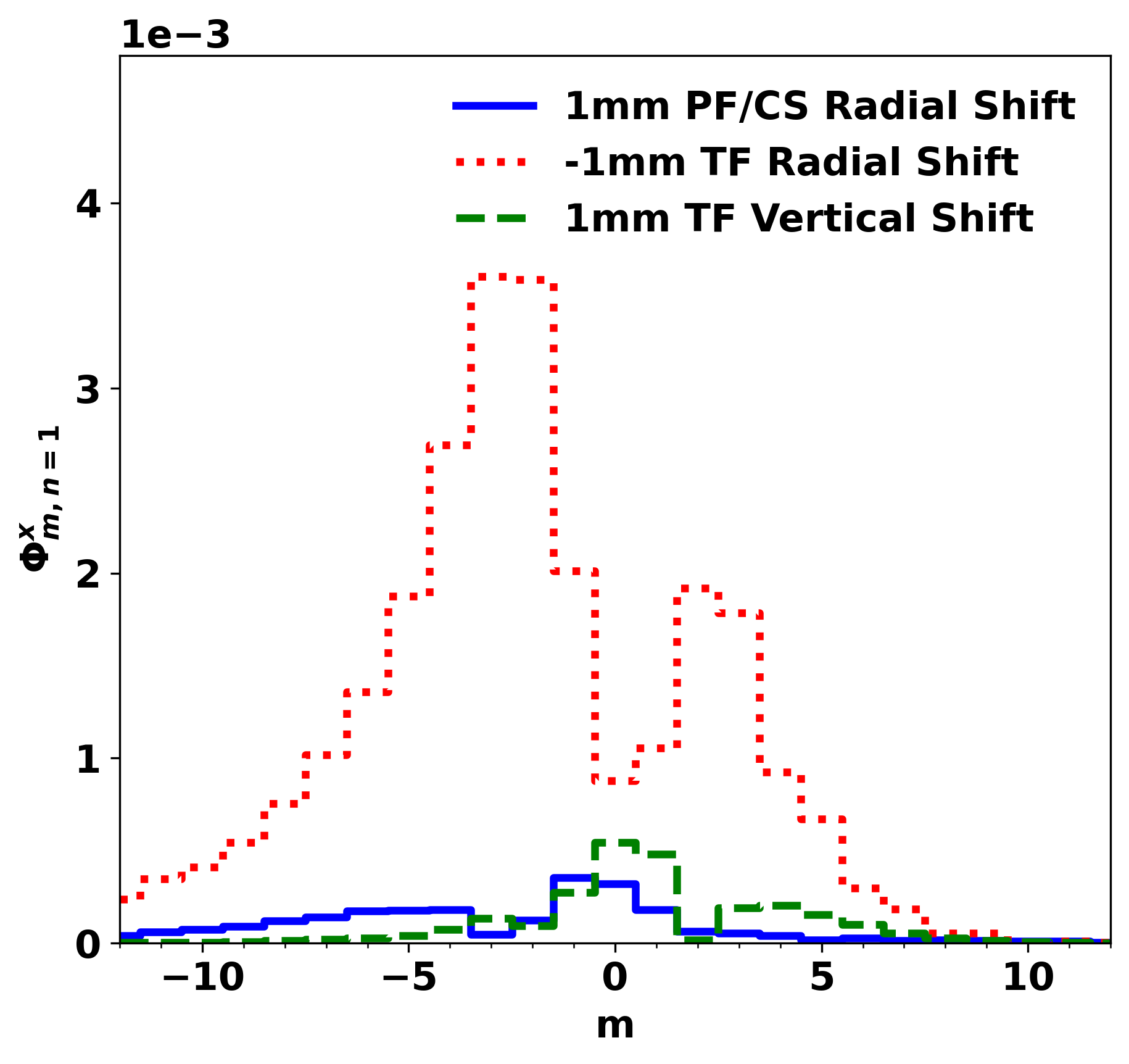}
    \caption{Energy normalized applied flux spectrum, $\Phi^x_m$, versus poloidal mode number in Hamada coordinates, $m$, for both cases from \cref{fig:ref_frame}(b) as well as a 1mm TF vertical shift. \label{fig:appliedspectrum_vertical}}
\end{figure}

To resolve this discrepancy, we include the vertical TF spectrum with the previously considered radial PF/CS and TF applied flux spectra in \cref{fig:appliedspectrum_vertical}. It is clear that although the maximum amplitudes of the normal field are similar, the vertical shift of the TF coils produces higher amplitude components at lower poloidal mode number, which we intuitively expect to couple more strongly to the magnetic axis. This analysis indicates that the reference frame can be approximately determined by high magnitude, low poloidal mode number components of the applied non-axisymmetric field spectrum. However, if the components are not sufficiently large, the reference frame will not be fixed with respect to that coil set, as in the case of the vertical TF shift.

This leads to a rather counterintuitive result: while the vertical TF shifts produce maximum perturbed normal fields almost an order of magnitude smaller than an equivalent radial TF shift, perturbative models will ultimately retain validity for a larger range of radial shifts than vertical shifts. This is due to the magnetic axis being fixed to the radial centroid of the TF coils, allowing the user to approximately know the perturbed axis location without the use of a nonlinear model and to move to a frame where the perturbed axis is axisymmetric. This avoids the requirement from the linearized MHD formulation that the perturbation to the axis is small. 

\section{\label{sec:Linearity} Linearity of Error Fields from Poloidal Field and Central Solenoid Coil Shifts}

Even in the correct reference frame, perturbative codes are still limited by the underlying linearized MHD theory. In this section, we use the nonlinear equilibria calculated by VMEC to locate where this theory begins to lose validity.

\subsection{Amplitude of Displacement vs. Toroidal Mode Number}

We begin by analyzing the normal displacement of magnetic field lines, $\vec{\xi} \cdot \hat{n}$, in the nonlinear equilibria. This is an important quantity in perturbative models and is related to the perturbed magnetic field through
\begin{equation}
    \delta \vec{B} = \nabla \times \left({\vec\xi \times \vec{B}_0}\right).
\end{equation}
We approximate the normal displacement in VMEC  by evaluating the difference in location of a magnetic field line between the 3D shifted (subscript $s$) and reference (subscript $r$) equilibrium, 
\begin{equation}
    \vec{\xi}\cdot \hat{n} = \left(\vec{r}_s - \vec{r}_{r}\right) \cdot \hat{n}_{r},
    \label{eq:xi_n}
\end{equation}
where in this analysis the reference equilibrium is the SPARC equilibrium from the unshifted coil set. 
We calculate \cref{eq:xi_n} in PEST coordinates $\left(s, \theta^*, \zeta\right)$, where $s$ is the radial-like coordinate, $\theta^*$ the straight-fieldline poloidal angle, and $\zeta=\phi$ the cylindrical toroidal angle \cite{hirshmanSteepestdescentMomentMethod1983}. 

\begin{figure}
    \centering
    \includegraphics[width=1\linewidth]{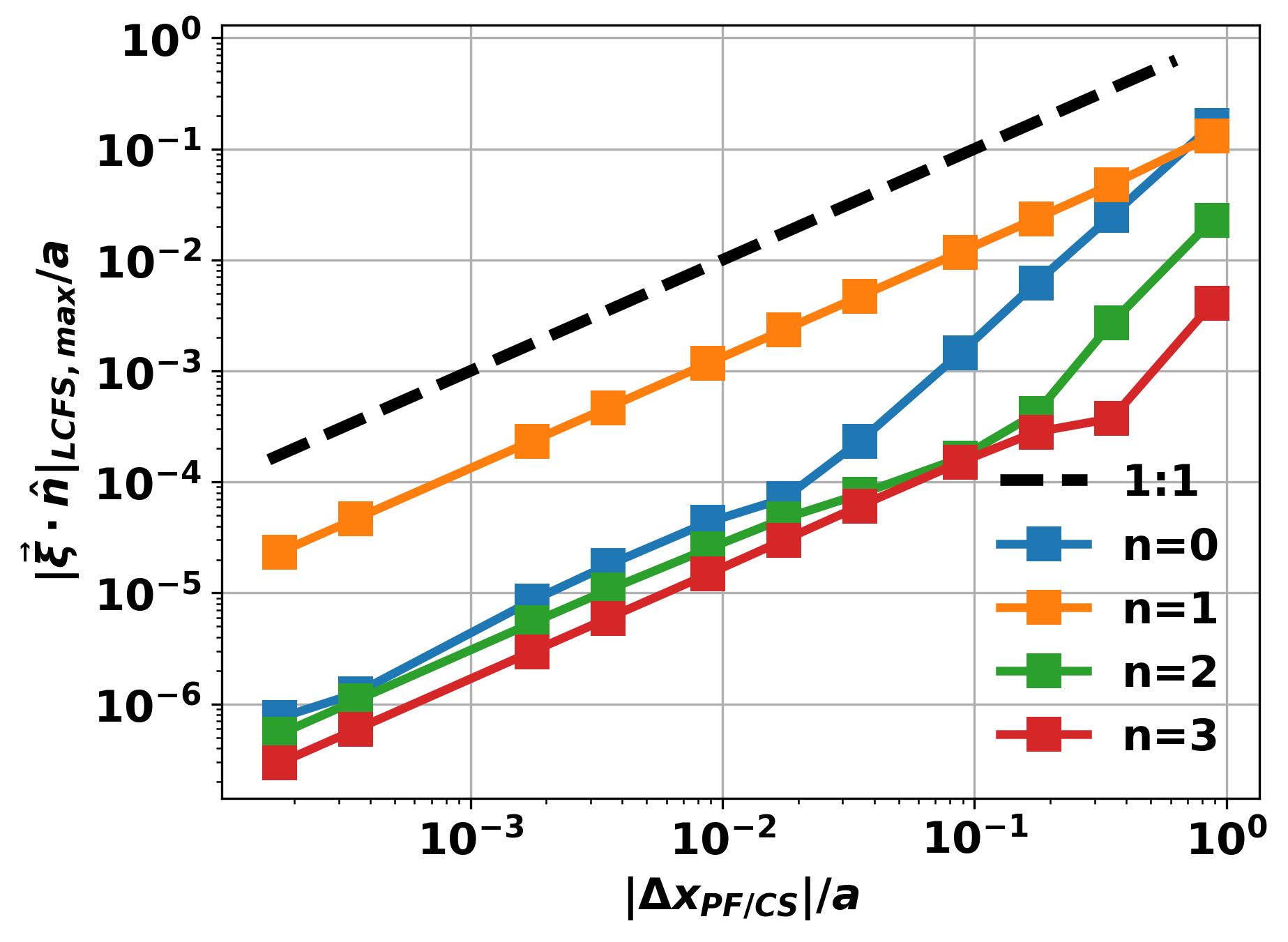}
    \caption{Maximum normal displacement amplitude from VMEC equilibria at the last closed flux surface (LCFS) for several toroidal mode numbers ($n$) as a function of PF/CS coil set shift. Both quantities are normalized to the minor radius of the equilibrium.}
    \label{fig:multin_PFCS}
\end{figure}

As we have shown previously that the TF centroid is the correct frame for the radial shifts of the SPARC equilibrium coils, we will consider shifts of the PF/CS coil set only. \Cref{fig:multin_PFCS} shows the maximum amplitude (poloidally) of the normal displacement at the last closed flux surface (LCFS) for several toroidal mode numbers versus the shift in the PF/CS coil set. Both quantities are normalized to the minor radius $a$ and again we plot a 1:1 line, with a shift corresponding to an equal normal displacement of the LCFS. 

At small shifts less than $1 \%$ of the minor radius, the $n=1$ response dominates by over an order of magnitude above other modes and remains approximately linear over the entire range of shifts plotted. As the level of asymmetry increases, other toroidal mode numbers become non-negligible, most notably $n=0$, becoming larger than the $n=1$ amplitude for shifts on the order of the minor radius. The inclusion of multiple mode numbers does not inherently invalidate the perturbative model, as toroidal mode numbers are decoupled in perturbed equilibria and the response can be simply summed across multiple modes. However, the $n\neq1 $ amplitudes would still linearly increase with shift amplitude if the perturbations were in the linear regime, and they clearly become nonlinear in these equilibria. Furthermore, perturbative models cannot calculate $n=0$ displacements, which is the largest $n\neq1 $ term.

With \cref{fig:multin_PFCS} in mind, we would like to create a metric to evaluate the importance of nonlinear effects as a function of shift. Instead of plotting each toroidal mode number amplitude individually, we instead take the maximum displacement at each toroidal angle from VMEC, $\vec{\xi}_{\max} = \max(|\vec{\xi}\cdot\hat{n}|)|_{\phi}$, to incorporate the multi-mode, nonlinear effects. We then define the difference in the linear vs. nonlinear prediction for this parameter at a shift $\Delta$ as 
\begin{equation}
    \vec{\xi}_{\max,\mathrm{diff}} =\vec{\xi}_{\max}(\Delta) - \frac{\Delta}{\Delta_{ref}}\vec{\xi}_{\max}(\Delta_{ref}),
\end{equation}
where $\Delta_{ref}$ is a small shift well in the linear regime (we used $0.5 ~\mathrm{mm}$). In other words, we calculate the difference in $\vec{\xi}_{\max}$ determined in the nonlinear equilibrium and the linear scaling that would be predicted from a perturbative model. This allows us to identify where small deviations from linear theory begin without directly comparing the outputs from perturbative codes, which do not agree exactly with VMEC due to how each code treats the response at rational surfaces \cite{reimanTokamakPlasmaHigh2015, kingExperimentalTestsLinear2015}.

\begin{figure*}
    \centering
    \includegraphics[width=1\linewidth]{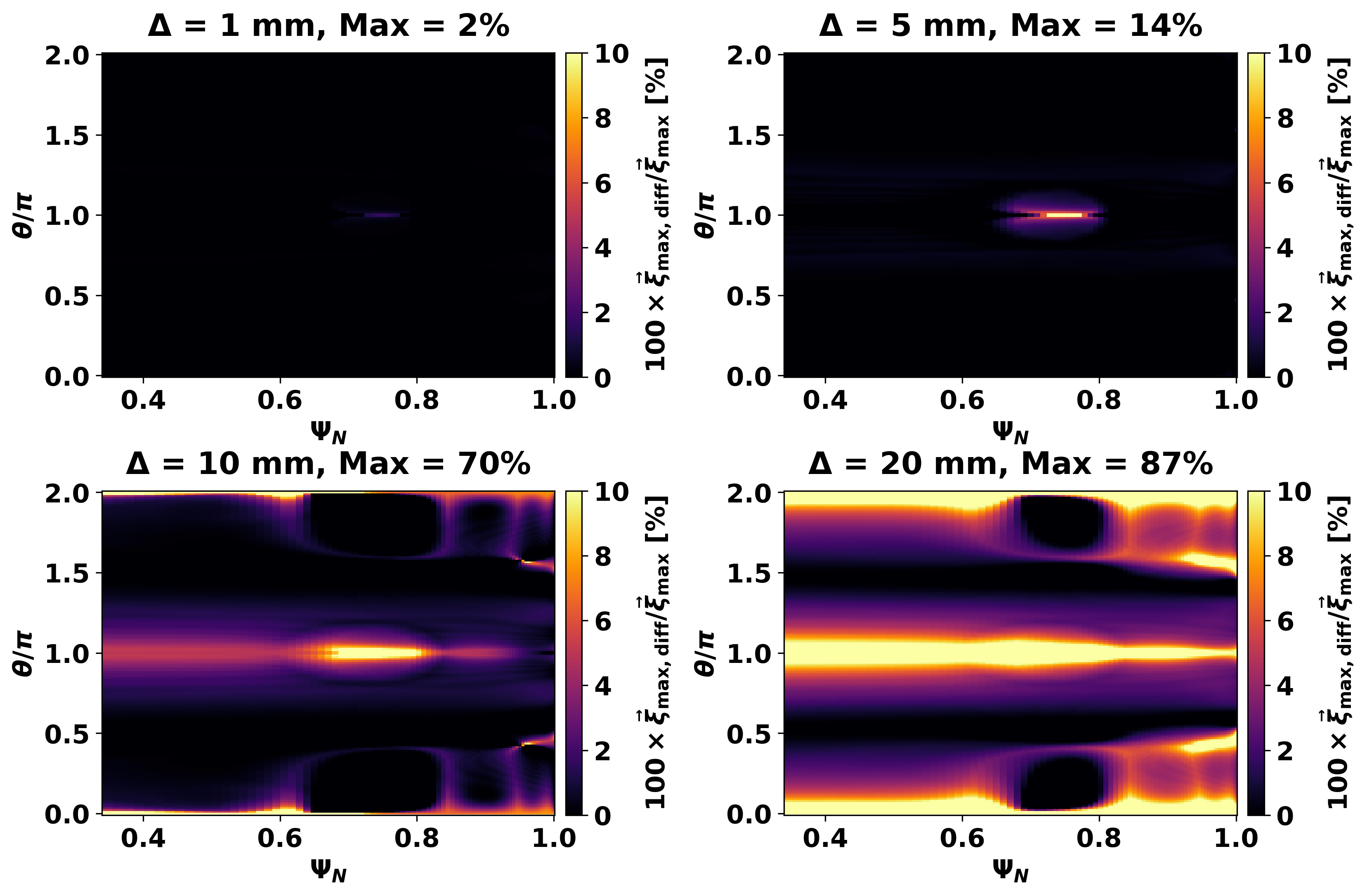}
    \caption{Percent difference between the nonlinear and linear prediction of the maximum displacement along a constant toroidal angle from VMEC equilibria as a function of both normalized poloidal flux and poloidal angle for four shifts of the PF/CS coils: $1, 5, 10, $ and $20 ~\mathrm{mm}$. The outboard and inboard midplane are at $\theta/\pi=0$ and $\theta/\pi=1$, respectively.}
    \label{fig:displacement_2D}
\end{figure*}

We plot the percent difference of $\vec{\xi}_{\max,\mathrm{diff}}$ as a function of the poloidal angle and poloidal flux $\Psi_N$ for several shifts in \cref{fig:displacement_2D}. Note that we have set the colorbar to range from $0$ to $10\%$ for all subplots, with all values $>10\%$ the same color. When normalized to the SPARC minor radius of $0.57~\mathrm{m}$, these shifts of $1~\mathrm{mm}$ to $20~\mathrm{mm}$ correspond to approximately $1.7\times10^{-3}$ to $3.4\times10^{-2}$ in the normalized shift units of \cref{fig:multin_PFCS}. For small shifts, i.e. $1$ and $5\mathrm{mm}$, a large majority of the response can be predicted by linear theory, with maximum differences of around $14\%$ on the inboard side. By $10~\mathrm{mm}$, a larger area has become invalid with maximum discrepancies between nonlinear findings and linear predictions of over $70\%$, and differs by near $90\%$ and in most of the plasma by $20~\mathrm{mm}$. The discrepancies are primarily localized to the outboard and inboard midplanes at $\theta = 0$ and $\theta = 1$, respectively. These findings match our expectations based on the relative amplitude for $n=1$ and other toroidal mode numbers in \cref{fig:multin_PFCS}, as this is where the $n=1$ and $n=0$ components begin to differ by less than a full order of magnitude.

\begin{figure*}
    \centering
    \includegraphics[width=1.0\linewidth]{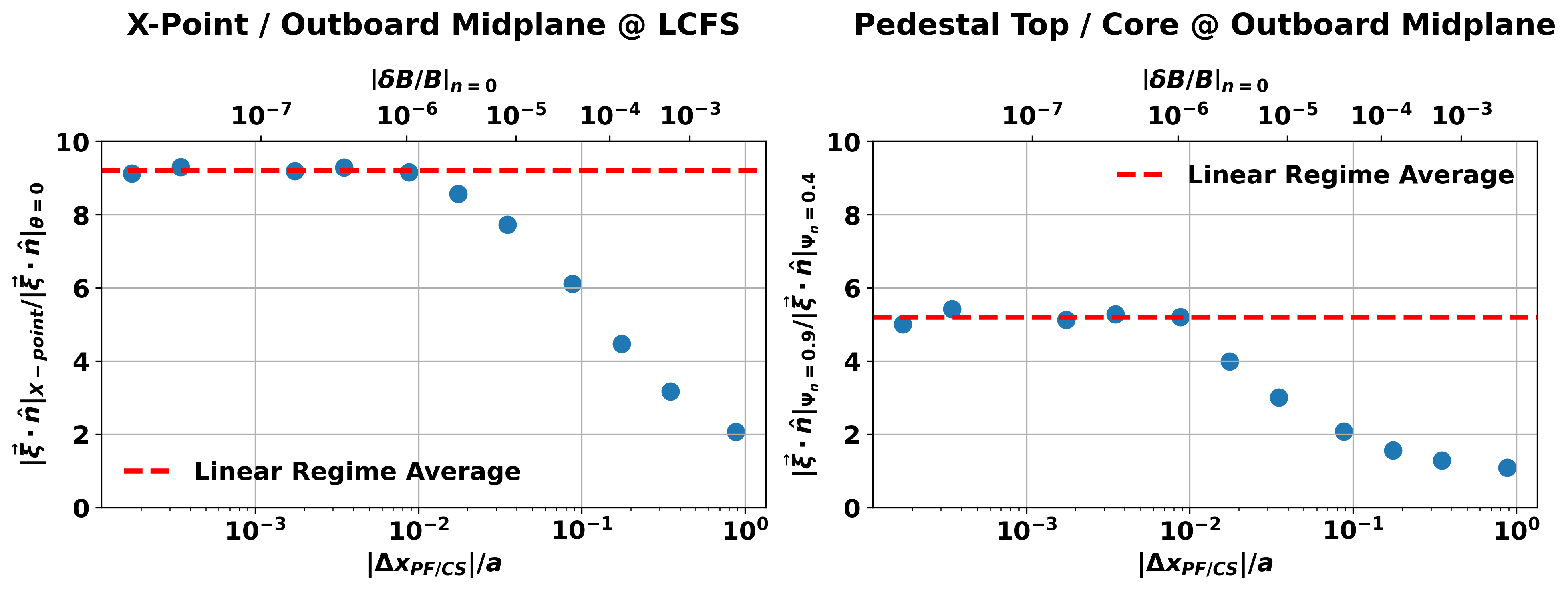}
    \caption{Ratio of the maximum normal displacement along a toroidal angle from VMEC equilibria at the X-point/outboard midplane displacement of the LCFS and pedestal top/core at the outboard midplane. We plot the ratios as a function of both the PF/CS shift normalized to the minor radius and the $n=0$ component of $\delta B / B$ at the magnetic axis. We plot the average of the linear regime data points as a red dashed line to emphasize when the ratio changes. 
    \label{fig:displacement_ratios}}
\end{figure*}

To determine more directly when the linear approximations begin to break down, we distill this maximum displacement toroidally into a 0D metric for the radial and poloidal structure as a function of shift. For the radial structure, we take the ratio between the pedestal top ($\Psi_N = 0.9$) and core ($\Psi_N = 0.4$) at the outboard midplane. For the poloidal structure, we use the ratio between the X-point (approximated by the poloidal max) and outboard midplane ($\theta=0$) at the LCFS. These ratios are commonly used to interpret the outputs of perturbative codes in terms of the plasma kink and peeling response \cite{liuModellingPlasmaResponse2011, liuELMControlRMP2016, liModellingPlasmaResponse2016}, and should be constant in regions where linear physics dominate. We plot the ratios in \cref{fig:displacement_ratios} versus both shift and the equivalent $n=0$ component of $\delta B / B$ at the magnetic axis, which we expect to be the largest contributor to breaking linearity based on \cref{fig:multin_PFCS}. There is a clear turning point from a linear to nonlinear response in both the radial and poloidal structure for $\delta B / B \approx 10^{-6}$ and shifts of $1\%$ of the minor radius; in non-normalized units, this corresponds to PF/CS shifts of around $5~\mathrm{mm}$ for the SPARC example considered here. Ultimately, this is slightly above tolerances in modern tokamaks, which are typically a few millimeters \cite{wengeAlignmentAssemblyEAST2005, bakStatusKSTARTokamak2006, shibanumaAssemblyStudyJT60SA2013}, validating the use of perturbative MHD models for setting these tolerances. However, this is nearing the limit identified here, which implies nonlinear modeling can become necessary for devices with larger relative tolerances or construction asymmetries. For example DIII-D possesses a $5~\mathrm{mm}$ shift of the inner TF legs \cite{maslineMisalignmentMagneticField2022}, which corresponds to $7\times 10^{-3}$ in the normalized units of \cref{fig:displacement_ratios} and is right on the cusp of the nonlinear regime, especially when considering it also has a tilt. NSTX-U, as built in 2016, also had a $\approx5~\mathrm{mm}$ centerpost shift in addition to its large tilt \cite{ferraroErrorFieldImpact2019}. This implies that if construction is not excellent, nonlinear modeling might become necessary to accurately capture the plasma response to error fields. 

\section{\label{sec:Summary} Summary and Future Work}

Perturbative MHD models are the most common choice for determining coil manufacturing and assembly tolerances for tokamaks. The linearized theory is a powerful and efficient method for determining the plasma response to small asymmetries in the applied field; however, it must be used correctly and within its limits to accurately predict the relevant physics. In this study, we have addressed two of these limitations using fully 3D, nonlinear equilibria generated from VMEC, an equilibrium solver primarily used in the stellarator field. The first limitation deals with the choice of which reference frame to perturb the axisymmetric plasma from, which strongly affects the applied spectrum on the plasma surface and therefore the plasma response. We have demonstrated through various coil set shifts on both a SPARC and NSTX-U equilibrium that the magnetic axis location is strongly dependent on the toroidal field coils, in particular the radial location of the inner legs at the midplane, which sets the reference frame. However, there was no clear reference frame for vertical shifts of the toroidal field coils, indicating more complex associations between the applied field spectrum and the axis location. We have shown that using the correct reference frame for the input coil sets in perturbative models is essential for obtaining the correct plasma response. 

We have also addressed the limitation of the linearized theory by determining where the magnetic field line displacement begins to demonstrate nonlinear effects. We find that, for our SPARC equilibrium example, the response is linear to around $5~\mathrm{mm}$ radial shifts of the poloidal field and central solenoid coil sets, validating the use of perturbative models to set tolerances on SPARC \cite{loganErrorFieldPhysics2023}. However, this value is sufficiently small to be notable for current devices with larger error fields such as DIII-D or NSTX-U (as built in 2016) or for future devices that are designed to be less sensitive to error fields and increased relative tolerances. 

Furthermore, we only considered coil set asymmetries which introduced dominantly $n=1$ fields on the plasma. In addition to being the most harmful to confinement, $n=1$ fields are also unique in that we can recover an axisymmetric equilibrium by modifying the reference frame, as we have shown above. This feature allows us to extend the validity of perturbative models by decomposing the field into the component which produces the error field and that which shifts the axis of symmetry, translating the coil set in space to use only the former as the input 3D field for the model. There is no equivalent reference frame for $n\geq2$ error fields, as axisymmetry cannot be recovered by shifting to a new frame. As a result, perturbative models are valid only when error fields are sufficiently small to keep the equilibrium axis unchanged and displacements in the linear regime. This limits their applicability to study error fields from TF coils, which we have shown to have a stronger effect on the magnetic axis and tend to be more prone to produce higher toroidal mode numbers due to the narrow toroidal extent of each coil. 

Ultimately, this study only analyzed two coil sets. While many quantities are normalized when possible, more work would be needed to make confident generalizations to the broader tokamak field. However, this workflow could be used to analyze any specific shift, i.e. identify from nonlinear equilibria how the axis moves for a given asymmetry, and run the perturbative model in this frame. For example, if we were specifically interested in the vertical shifts from \cref{sec:GeneralRefFrames}, we could have tilted the frame to the tilt of the magnetic axis from \cref{fig:n1_vertical}.
Furthermore, we only considered the plasma response through the normal displacement of magnetic field lines. 
One of the effects of error fields is through driving magnetic islands at rational surfaces, where VMEC is known to have difficulties converging \cite{lazersonVerificationIdealMagnetohydrodynamic2016}. 
Extensions of this study could incorporate other stellarator models such as DESC \cite{dudtDESCStellaratorEquilibrium2020} or PIES/CAS3D \cite{nuhrenbergMagneticSurfaceQualityNonaxisymmetric2009} to directly predict the linearity of the island width and other confinement metrics such as neoclassical toroidal viscosity. 

\section{\label{sec:Acknowldgements}Acknowledgments}

This material is based upon work supported by the U.S. Department of Energy, Office of Science, Office of Advanced Scientific Computing Research, Department of Energy Computational Science Graduate Fellowship under Award Number DE-SC0024386. The work was also supported by the U.S. Department of Energy Office of Science Office of Fusion Energy Sciences under Award DE-SC0022272.
We also thank Ian Stewart for generating the NSTX-U equilibrium used in this work, and Commonwealth Fusion Systems for coil and equilibrium data for the SPARC tokamak. 

\begin{appendix}
\section{\label{app:coils}Generation of Coil Sets}

We implemented the coil shifts within the Simsopt  framework \cite{landremanSIMSOPTFlexibleFramework2021} to leverage its existing interface with VMEC; however, any of the coil configurations here can be generalized to any coil representation.

The toroidal field coils are described as Fourier series in Cartesian coordinates,
\begin{equation}
    \vec{x} = \vec{x}_0 + \sum_{n=1}^{N} \left(\vec{x}_{c,n} \cos(nt) + \vec{x}_{s,n}\sin(nt)\right),
    \label{eq:XYZFourier}
\end{equation}
where $\vec{x} = (x, y, z)$, $\vec{x}_0$ is the coil centroid, $N$ is the number of Fourier modes included in each coil, and $t\in[0, 2\pi)$. Each physical coil is composed of two filaments parameterizing the inner and outer cross section The Fourier coefficients are determined from a set of Cartesian points describing the nominal coil set. We obtain a $n=1$ radial shift $\Delta x$ by moving all coils uniformly the $\hat{x}$ direction, $x_0 \rightarrow x_0 + \Delta x$. Similarly, a $n=1$ vertical shift $\Delta z$ is obtained similarly from $z_0 \rightarrow z_0 + z_0 \cos(\phi)$, where $\phi = \arctan(y_0 / x_0)$.

For a tilt of the centerpost about the bottom of the device, we first separate the unshifted coil data points into the inner legs, defined as the set of all points whose radial position $R = \sqrt{x^2 + y^2}$ is less than the minimum for the coil, $R_{min}$ (with a small buffer of around $1~\mathrm{cm}$). We then define the outer legs as the set of all points with radial position greater than $R_{min} + 2 Z_{max} \sin(\theta)$ (again with a buffer), i.e. all points outside the radius where the inner legs are located after tilting. Here, $Z_{max}$ is half the height of the machine, and $\theta$ is the tilt angle. We remove all points with $R_{min} < R < R_{min} + 2 Z_{max} \sin(\theta)$. We then tilt the centerpost by first shifting it vertically by $Z_{max}$ to place the pivot point at the origin, using rotation packages within $\texttt{scipy}$ to align the inner leg points with the tilt unit vector, $(\cos \theta, 0, \sin \theta)$, and restoring the pivot point back to its initial location by shifting by $-Z_{max}$. We then combine the shifted inner legs and unshifted outer legs and obtain a new Fourier representation according to \cref{eq:XYZFourier}. We use a high number of Fourier modes $N$ ($\approx 50$) to reduce ringing artifacts produced from the Fourier fit. The process to define a centerpost shift is identical, except with the rotation replaced by a shift in $+\hat{x}$.

The poloidal field coils and central solenoid are represented as planar curves using Fourier series in polar coordinates, 
\begin{equation}
    r(\phi) = r_0 + \sum_{n=1}^{N} r_{c,n} \cos(n\phi) + r_{s,n} \sin(n\phi),
\end{equation}
with the coil centroid at $(x_0, y_0, z_0)$ and rotation defined by the quaternion $[a, b, c, d]$. We uniformly distribute the filaments in a rectangle in $R,Z$ which define each coil by setting $r_0$ and $z_0$, with the amount determined by a convergence scan in the applied field. We can then obtain a radial shift $\Delta x$ through $x_0 \rightarrow x_0 + \Delta x$, and a vertical tilt by setting the quaternion to align the coil normal vector with the tilt unit vector, described above. 

For the centerpost tilt, we only modify the CS and PF1 coils, leaving PF2-5 unshifted. For a tilt $\theta$, we first shift the coil in space through $x_0 \rightarrow (z_0 - Z_{max})\sin(\theta)$ and $z_0 \rightarrow (z_0 - Z_{max})\cos(\theta) + Z_{max}$. We then rotate the coil in place by aligning the coil normal with the tilt vector using the appropriate quaternion. 

\end{appendix}

\printbibliography

\end{document}